\documentclass[pdflatex,sn-mathphys-num,iicol,a4paper]{sn-jnl}

\usepackage{graphicx}%
\usepackage{multirow}%
\usepackage{amsmath,amssymb,amsfonts}%
\usepackage{amsthm}%
\usepackage{mathrsfs}%
\usepackage[title]{appendix}%
\usepackage{xcolor}%
\usepackage{textcomp}%
\usepackage{manyfoot}%
\usepackage{booktabs}%
\usepackage{algorithm}%
\usepackage{algorithmicx}%
\usepackage{algpseudocode}%
\usepackage{listings}%
\usepackage{dcolumn}%
\usepackage{bm}%
\usepackage{siunitx}

\usepackage{booktabs}
\usepackage{soul}
\usepackage[utf8]{inputenc}
\usepackage{hyperref}
\usepackage{caption}
\usepackage{stfloats}
\usepackage{placeins}

\DeclareSIUnit\electroncharge{e}
\DeclareSIUnit\atomicmassunit{u}

\begin{document}


\title[Mass measurements of \textsuperscript{179--184}Yb identify an anomalous proton-neutron interaction]{Mass measurements of \textsuperscript{179--184}Yb identify an anomalous proton-neutron interaction}


\author[1]{\fnm{C.~L.} \sur{Brown}}
\author[2]{\fnm{J.} \sur{Ash}}
\author[3]{\fnm{B.} \sur{Ashrafkhani}}
\author[4]{\fnm{J.} \sur{Bergmann}}
\author[5]{\fnm{T.} \sur{Brunner}}
\author[2,6]{\fnm{J.~D.} \sur{Cardona}}
\author[7,8]{\fnm{R.~B.} \sur{Cakirli}}
\author[9]{\fnm{R.~F.} \sur{Casten}}
\author[2]{\fnm{C.} \sur{Chambers}}
\author[4,10]{\fnm{T.} \sur{Dickel}}
\author[6]{\fnm{G.} \sur{Gwinner}}
\author[2,5]{\fnm{Z.} \sur{Hockenbery}}
\author[2,13]{\fnm{A.} \sur{Jacobs}}
\author[2,6]{\fnm{J.} \sur{Lassen}}
\author[2,14]{\fnm{R.} \sur{Li}}
\author[11]{\fnm{D.} \sur{Lunney}}
\author[2,6]{\fnm{S.} \sur{Kakkar}}
\author[2]{\fnm{F.} \sur{Maldonado~Mill\'{a}n}}
\author[12,7]{\fnm{N.} \sur{Minkov}}
\author[2,4]{\fnm{A.} \sur{Mollaebrahimi}}
\author[2,13]{\fnm{E.~M.} \sur{Lykiardopoulou}}
\author[2]{\fnm{S.} \sur{Paul}}
\author[4,10]{\fnm{W.~R.} \sur{Pla{\ss}}}
\author[2]{\fnm{W.~S.} \sur{Porter}}
\author[2]{\fnm{D.} \sur{Ray}}
\author[1]{\fnm{M.~P.} \sur{Reiter}}
\author[2]{\fnm{A.} \sur{Ridley}}
\author[4,10,16]{\fnm{C.} \sur{Scheidenberger}}
\author[2]{\fnm{R.} \sur{Simpson}}
\author[2,6]{\fnm{C.} \sur{Walls}}
\author[2,13]{\fnm{Y.} \sur{Wang}}
\author[2]{\fnm{A.~P.} \sur{Weaver}}
\author[2,15]{\fnm{A.~A.} \sur{Kwiatkowski}}


\affil[1]{\orgdiv{School of Physics and Astronomy}, \orgname{University of Edinburgh}, \orgaddress{\city{Edinburgh}, \state{Scotland}, \postcode{EH9 3FD}, \country{United Kingdom}}}
\affil[2]{\orgname{TRIUMF}, \orgaddress{\street{4004 Wesbrook Mall}, \city{Vancouver}, \state{British Columbia}, \postcode{V6T 2A3}, \country{Canada}}}
\affil[3]{\orgdiv{Department of Physics and Astronomy}, \orgname{University of Calgary}, \orgaddress{\street{2500 University Drive NW}, \city{Calgary}, \state{Alberta}, \postcode{T2N 1N4}, \country{Canada}}}
\affil[4]{\orgdiv{II. Physikalisches Institut}, \orgname{Justus-Liebig-Universit{\"a}t}, \orgaddress{\postcode{35392}, \city{Gie{\ss}en}, \country{Germany}}}
\affil[5]{\orgdiv{Department of Physics and Astronomy}, \orgname{McGill University}, \orgaddress{\street{3600 Rue University}, \city{Montr{\'e}al}, \state{Quebec}, \postcode{H3A 2T8}, \country{Canada}}}
\affil[6]{\orgdiv{Department of Physics and Astronomy}, \orgname{University of Manitoba}, \orgaddress{\city{Winnipeg}, \state{Manitoba}, \postcode{R3T 2N2}, \country{Canada}}}
\affil[7]{\orgname{Max-Planck-Institut f{\"u}r Kernphysik}, \orgaddress{\street{Saupfercheckweg 1}, \city{Heidelberg}, \postcode{69117}, \country{Germany}}}
\affil[8]{\orgname{ExtreMe Matter Institute EMMI, GSI Helmholtzzentrum f{\"u}r Schwerionenforschung GmbH}, \orgaddress{\street{Planckstra{\ss}e 1}, \city{Darmstadt}, \postcode{64291}, \country{Germany}}}
\affil[9]{\orgname{Wright Lab, Yale University}, \orgaddress{\city{New Haven}, \state{Connecticut}, \postcode{06520}, \country{USA}}}
\affil[10]{\orgname{GSI Helmholtzzentrum f{\"u}r Schwerionenforschung GmbH}, \orgaddress{\street{Planckstra{\ss}e 1}, \city{Darmstadt}, \postcode{64291}, \country{Germany}}}
\affil[11]{\orgname{Universit{\'e} Paris-Saclay, CNRS/IN2P3, IJCLab}, \orgaddress{\postcode{91405}, \city{Orsay}, \country{France}}}
\affil[12]{\orgname{Institute for Nuclear Research and Nuclear Energy, Bulgarian Academy of Sciences}, \orgaddress{\street{Tzarigrad Road 72}, \postcode{BG-1784}, \city{Sofia}, \country{Bulgaria}}}
\affil[13]{\orgdiv{Department of Physics \& Astronomy}, \orgname{University of British Columbia}, \orgaddress{\city{Vancouver}, \state{British Columbia}, \postcode{V6T 1Z1}, \country{Canada}}}
\affil[14]{\orgdiv{Department of Physics}, \orgname{University of Windsor}, \orgaddress{\city{Windsor}, \state{Ontario}, \postcode{N9B 3P4}, \country{Canada}}}
\affil[15]{\orgdiv{Department of Physics \& Astronomy}, \orgname{University of Victoria}, \orgaddress{\street{3800 Finnerty Road}, \city{Victoria}, \state{British Columbia}, \postcode{V8P 5C2}, \country{Canada}}}
\affil[16]{\orgname{Helmholtz Research Academy Hesse for FAIR (HFHF), GSI Helmholtz Center for Heavy Ion Research}, \orgaddress{\city{Gie{\ss}en}, \postcode{35392}, \country{Germany}}}

\abstract{Mass measurements of nuclei can identify structurally-driven trends in binding energy across isotopic chains, and can also isolate specific nucleon-nucleon interactions, such as the $\delta V_{\mathrm{pn}}$ interaction of the last two valence protons with the last two valence neutrons. Below \textsuperscript{208}Pb, investigation of the local binding energy and $\delta V_{\mathrm{pn}}$ systematics can facilitate a better understanding of the behaviour of the proton-neutron interaction in the `hole-hole' regime (where valence interactions can be modelled in hole-space rather than particle-space) and provide insight on the potential onset of a prolate-to-oblate shape transition. However, measurement of the necessary nuclei has been exceptionally challenging. Here we present six first-time measurements of neutron-rich ytterbium, using advanced rare isotope production and mass spectrometry techniques, leading to the identification of an anomalously strong proton-neutron interaction in the `hole-hole' quadrant below \textsuperscript{208}Pb. The scale of this interaction, at \textsuperscript{186}Hf, is comparable to that of similar signals at doubly-magic nuclei and shape transitions. The experimental results are compared with contemporary mean-field model predictions, that do not accurately reproduce the anomaly. The results are also used to benchmark predictions from several models, to facilitate more accurate descriptions towards a key \textit{r}-process waiting point at $N=126$.}

\maketitle


\section{Main}

\begin{figure*}[!b]
    \centering
    \includegraphics[width=0.8\linewidth]{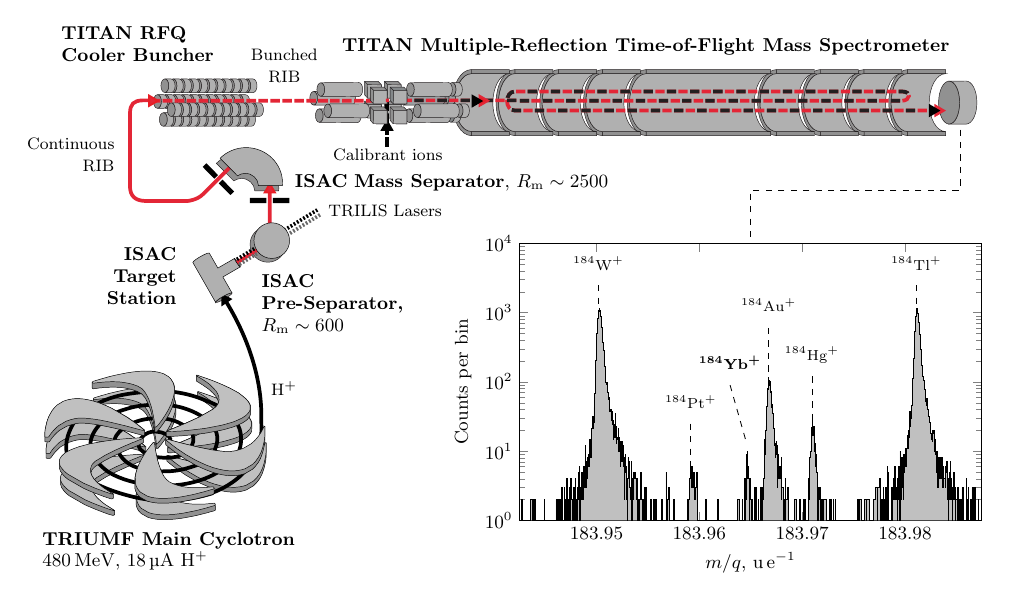}
    \caption{\textit{Schematic of TITAN and the upstream ISAC beamline at TRIUMF, not to scale. An intense \SI{480}{\mega\electronvolt}, \SI{18}{\micro\A} H\textsuperscript{+} beam was provided by the TRIUMF main cyclotron, and impinged onto a thick UC\textsubscript{X} target to produce an array of radioactive species. These species were extracted and ionised to form a cocktail radioactive ion beam (RIB), with selective ionisation of ytterbium species achieved through application of TRIUMF's Resonant Ionisation Laser Ion Source (TRILIS). This cocktail RIB was twice mass separated and transported to TITAN's RFQ Cooler and Buncher, which then delivered the beam as radioactive ion bunches to TITAN's Multiple-Reflection Time-of-Flight Mass Spectrometer (MR-ToF-MS). Here narrow ion bunches, with calibrant ions injected from an internal source, were reflected for multiple turns between a set of two isochronous ion mirrors, facilitating precise mass measurement with a high mass resolving power. A mass spectrum for the reported results at $A=184$ is shown. The identified species are labelled, including \textsuperscript{184}Yb, the most exotic species measured. `Mass-selective re-trapping' centred on \textsuperscript{184}Yb was used to cut the rates of the major contaminant peaks by four orders of magnitude. See the main text for further details0}.}
    \label{fig:experiment}
\end{figure*}

The nuclear landscape features sudden structural changes, where the addition of only one or two nucleons results in a sudden macroscopic change in the ground-state shape of a nucleus~\cite{RevModPhys.83.1467,Casten2006}. This occurs in the rare-earth nuclei above the \textsuperscript{132}Sn closed-shell configuration, where the increasing number of valence protons and valence neutrons leads to a dominance of the deformation-favouring proton-neutron interaction over the sphericity-favouring pairing interaction~\cite{Federman1977,Federman1979}. Toward the `mid-shell' region, this leads to the spontaneous breaking of spherical symmetry~\cite{Iachello2000}, resulting in a critical spherical-to-prolate shape transition across $N=90$. At $N\sim116$, a prolate-to-oblate shape transition is anticipated ~\cite{PhysRevC.77.064322,Bonatsos2017}. Such a transition may have a significant impact on the astrophysical \textit{r}-process, which produces the heaviest elements in nature~\cite{Mumpower2016,Sun2008}. However, the matter has remained indeterminate due to difficulties in producing and measuring short-lived neutron-rich rare-earth isotopes. With contemporary nuclear models providing different predictions regarding this potential shape transition~\cite{Bonatsos2024}, this region has long represented an opportunity to improve our understanding of the emergent dynamics of the nuclear force.

Nuclear spectroscopy has been used to probe the structure of tungsten isotopes up to $N = 118$, providing indications of a gradual prolate-to-oblate transition~\cite{Sahin2024}, but for lighter elements similar measurements have only been obtained up to $N = 108$~\cite{PRITYCHENKO20161}. Without spectroscopy, mass measurements can offer insight, as trends in binding are key markers of underlying structural changes. Binding energy filters can isolate the strength of specific interactions, such as the $\delta V_{\mathrm{pn}}$ interaction between the final two protons and final two neutrons~\cite{Zhang1989}. Trends in these observables correlate with shape deformation~\cite{PhysRevLett.96.132501}, and can inform the development of new theoretical approaches~\cite{PhysRevC.88.054309,Bonatsos20172}. However, mass measurements are difficult for similar reasons. Very few first-time mass measurements have been secured in this region over the past three decades~\cite{Wang2021}.

Accordingly, this work was enabled by advances in the production and separation of neutron-rich rare-earth nuclei. Ytterbium isotopes were produced by TRIUMF's Isotope Separator and Accelerator (ISAC)~\cite{Dilling2014}, an isotope separation on-line (ISOL) facility. Protons were delivered to ISAC from the TRIUMF main cyclotron at \SI{480}{\mega\electronvolt} and \SI{18}{\micro\A}, as illustrated in Figure~\ref{fig:experiment}. Here protons were impinged on a contemporary uranium carbide (UC\textsubscript{X}) target structure~\cite{Cervantes2020}, at a newly accessible beam power of \SI{8.64}{\kilo\W}, producing a variety of exotic radioactive species. Following diffusion from the \SI{1950}{\celsius} target, ytterbium species were selectively ionised with TRIUMF's Resonant Ionisation Laser Ion Source (TRILIS)~\cite{Lassen2017}. A newly developed two-step resonant laser excitation scheme was applied~\cite{Lassen2017}, to achieve a factor-of-forty increase in the ionisation rate for ytterbium (compared to residual surface ionisation). This scheme, a variation of~\cite{Backe2007} that enables future applications in ion guide laser ion sources, drives an initial \SI{398.911}{\nano\m} transition and then a \SI{385.752}{\nano\m} transition to an auto-ionising state, with laser powers of \SI{285}{\milli\W} and \SI{330}{\milli\W} respectively. This step is crucial for ytterbium, as despite its low \SI{1470}{\celsius} boiling point (which allows for good diffusion and effusion from the target) it has a moderately high \SI{6.3}{\electronvolt} first ionisation energy that lowers surface ionisation efficiency. Following ionisation, ISAC's two-stage magnetic dipole mass separator selected subsequent mass numbers from $A=178$ to $A=184$, for delivery at \SI{20}{\kilo\electronvolt} to TRIUMF's Ion Trap for Atomic and Nuclear Science (TITAN)~\cite{Kwiatkowski2024}.

At TITAN ions were accumulated into bunches in the radiofrequency quadrupole (RFQ) cooler buncher, and then delivered to TITAN's Multiple Reflection Time-of-Flight Mass Spectrometer (MR-ToF-MS)~\cite{Reiter2021}. Within the MR-ToF-MS ions were re-cooled and transported through a series of RFQs and finally injected into the time-of-flight analyser. Here they completed a pre-determined number of turns between two electrostatic ion mirrors, with a total flight time of around \SI{20}{\milli\s}, until the downstream mirror was switched open to impinge ions onto a time-of-flight detector. This enabled direct mass determination, through calibration with known contaminants in the beam. Contamination was suppressed by a factor of $10^{4}$ through `mass-selective re-trapping'~\cite{Dickel2017}.

\setlength{\tabcolsep}{5.5pt}
\begin{table*}[!b]
  \centering
    \begin{tabular}{c c c c c c c}
    \hline
    \hline
    \\[-1em]
    ~~~~~Species & ~~Calibrant & Mass Ratio & ME$_{\text{TITAN}},$ $\text{keV}/c^{2}$~~& ME$_{\text{AME20}},$ $\text{keV}/c^{2}$~~& Difference, $\text{keV}/c^{2}$~~~~\\
    \\[-1em]
    \hline
    \\[-1em]
 ~~~~~$^{178}$Yb         &  ~~$^{140}$Ce$^{19}$F$_{2}$ & ~~1.000\,249\,74~(19)~~  & -49663~(31)~~ & -49677.1~(6.6)~~ &  ~14~(32)~~ \\
 ~~~~~$^{179}$Yb         &  ~~$^{160}$Dy$^{19}$F~      & ~~1.000\,146\,22~(16)~~  & -46791~(26)~~ & \#~~             & \#~~	  \\
 ~~~~~$^{180}$Yb         &  ~~$^{156}$Gd$^{12}$C$_{2}$ & ~~1.000\,165\,42~(15)~~  & -44811~(25)~~ & \#~~             & \#~~	  \\
 ~~~~~$^{181}$Yb         &  ~~$^{181}$Ta               & ~~1.000\,041\,17~(15)~~  & -41499~(26)~~ & \#~~             & \#~~	  \\
 ~~~~~$^{182}$Yb         &  ~~$^{163}$Dy$^{19}$F       & ~~1.000\,170\,05~(20)~~  & -39052~(34)~~ & \#~~             & \#~~	  \\
 ~~~~~$^{182}$Yb         &  ~~$^{182}$W                & ~~1.000\,054\,21~(25)~~  & -39058~(42)~~ & \#~~             & \#~~	  \\
 ~~~~~$^{183}$Yb         &  ~~$^{183}$W                & ~~1.000\,066\,02~(15)~~  & -35116~(26)~~ & \#~~             & \#~~	  \\
 ~~~~~$^{184}$Yb         &  ~~$^{184}$W                & ~~1.000\,078\,13~(22)~~  & -32318~(37)~~ & \#~~             & \#~~	  \\
 \hline
 \hline
\end{tabular}%
  \caption{\textit{Table of measurements. Calibrants, atomic mass ratios, and atomic mass excess results are given for each species and compared to the AME2020 experimental data~\cite{Wang2021}. Atomic mass ratios and atomic mass excesses (ME$_{\text{TITAN}}$, ME$_{\text{AME20}}$) are determined from the ionic values through inclusion of the electron mass, with all species measured as singly-charged ions. The atomic mass ratios are taken as the mass ratio of the ion-of-interest to the calibrant ion, $m_{\text{IoI}}/m_{\text{cal}}$. Previously unmeasured species are denoted with a \# symbol.}}
  \label{tab:mass_table}%
\end{table*}

Precision measurements were performed for \textsuperscript{178-184}Yb, with mass values extracted as described in the Methods and~\cite{Ayet2019}. Results are presented in Table~\ref{tab:mass_table}. Relative uncertainties of $\delta m/m \leq 2.2\cdot10^{-7}$ were assigned according to the same procedure, providing absolute uncertainties of $\delta m\leq$~\SI{37}{\kilo\electronvolt}$/c^{2}$. The \textsuperscript{178}Yb measurement is in agreement with the $\delta m =$~\SI{9.4}{\kilo\electronvolt} ISOLTRAP result~\cite{Huang2019}. All other measurements are first-time. These measurements extend the available mass data six neutrons further from the nearest stable ytterbium isotope, \textsuperscript{176}Yb, where previously data was only available up to \textsuperscript{178}Yb. For neutron-rich nuclei, mass data for ytterbium is now available further from stability than for any other element between $Z=65$ and $Z=75$, representing a significant step into the unknown and an unprecedented accomplishment for an MR-ToF system at an ISOL facility.

\begin{figure}[b]
    \includegraphics[width=\linewidth]{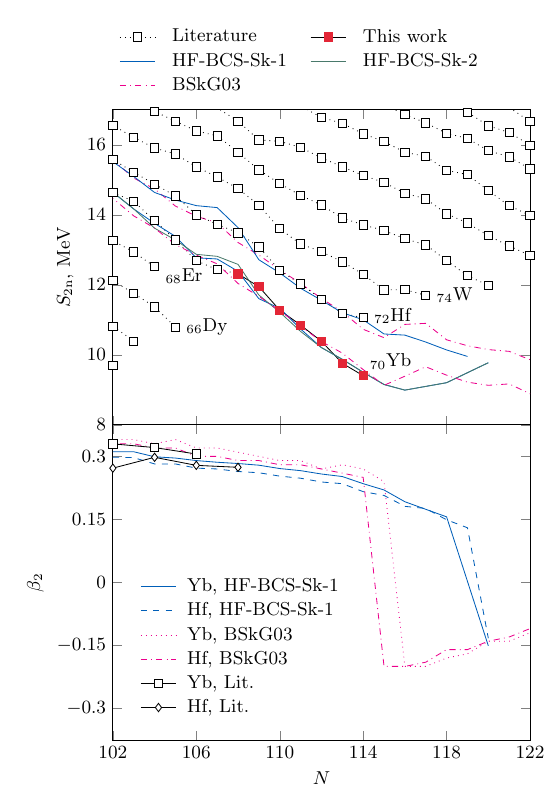}
    \caption{\textit{\textbf{Top}: Experimental two-neutron separation energies for the local even-$Z$ species, with data from the AME2020~\cite{Wang2021}, GSI~\cite{Shubina2013}, and this work. Theoretical predictions are overlaid for the \textsubscript{70}Yb and \textsubscript{72}Hf chains, from the BSkG03 model~\cite{Grams2023} and this work. Note that for the ytterbium $S_{2\mathrm{n}}$, after $N=115$, the HF-BCS-Sk-1 and HF-BCS-Sk-2 values are shown only for the even-$N$ ytterbium species. Red squares are the new ytterbium data. \textbf{Bottom}: Experimental $\beta_{2}$ deformations for \textsubscript{70}Yb and \textsubscript{72}Hf, with data from the NNDC~\cite{NNDC}. Theoretical predictions are overlaid, from the BSkG03~\cite{Grams2023} model and this work.  Note that for the ytterbium $\beta_{2}$, only the HF-BCS-Sk-1 values are shown, as the two sets of $\beta_{2}$ predictions are not distinguishable; after $N=115$ these values are shown for only the even-$N$ ytterbium species.}}
    \label{fig:s2n}
\end{figure}

We can investigate trends in binding through the two-neutron separation energies:
\begin{align}
S_{2\mathrm{n}}(N,Z)
&= E_{\mathrm{B}}(N,Z) 
 - E_{\mathrm{B}}(N-2,Z) \nonumber \\
&= \bigl\{ m(N-2,Z) + 2m_{\text{n}} \nonumber \\
&\quad - m(N,Z) \bigr\} \cdot c^{2}.
\end{align}
Here $m_\mathrm{n}$ is the neutron mass. The $S_{2\mathrm{n}}$ typically trend downwards as a function of $N$, as additional neutrons become incrementally less bound, with sharp drops occurring at `shell gaps.' Shape transitions are an exception, as the nuclei can become more bound due to a rapid increase in deformation energy, leading to a flat or positive $S_{2\mathrm{n}}$ gradient. The new data, shown in Figure~\ref{fig:s2n} (top), reveals that the flattening trend in ytterbium towards $N=108$ is followed by a drop from $N=108$ to $N=110$ that is similarly steep to the drop at the deformed $N=104$ sub-shell gap in ytterbium~\cite{Bengtsson1984}. This may be related to the presence of a deformed $N=108$ sub-shell gap, as in hafnium and tungsten~\cite{Barber1973,Watanabe2019}. However, this is not clear, as the steep gradient in ytterbium persists to $N=114$. It is notable that towards $N=114$ the trends in the hafnium and ytterbium chains diverge. Whereas $S_{2\mathrm{n}}$ falls rapidly from $N=112$ to $N=114$ in ytterbium, in hafnium there is a flattening trend. This indicates that the binding of the final neutrons is significantly higher in \textsuperscript{186}Hf, in relative terms, than in \textsuperscript{184}Yb. In hafnium, this flattening trend may indicate a change in structure~\cite{Shubina2013}. The same interpretation has been made of the flattening in tungsten towards $N=116$, an assessment corroborated by spectroscopic data~\cite{Mukai2025,Sahin2024}.

\begin{figure}
    \centering
    \includegraphics[width=\linewidth]{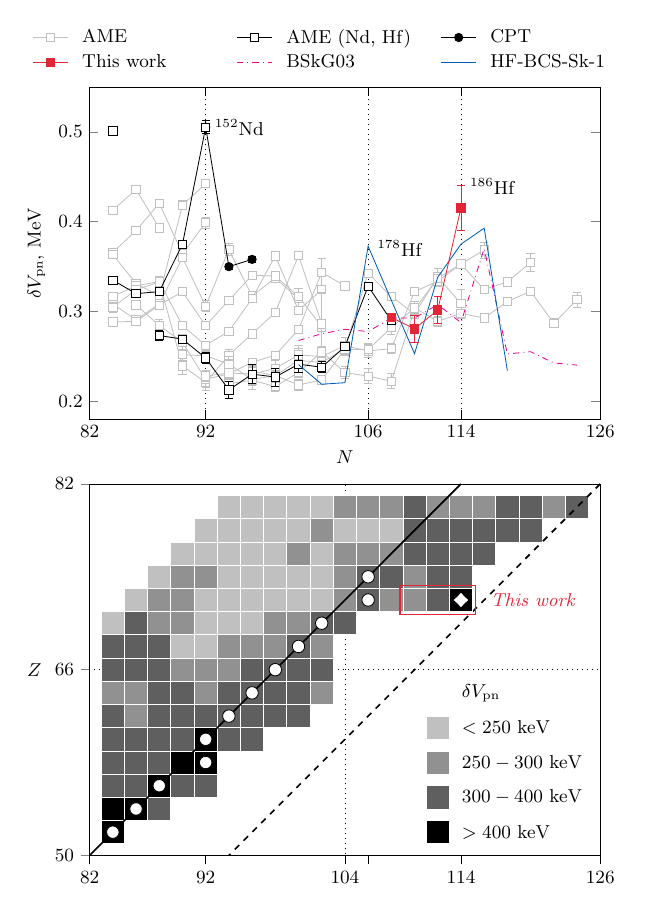}
    \caption{\textit{Experimental $\delta V_{\mathrm{pn}}$ values for the even-even species of the region between \textsuperscript{132}Sn and \textsuperscript{208}Pb, with data from the AME2020~\cite{Wang2021}, CPT~\cite{Orford2022}, and this work. \textbf{Top:} A 2-dimensional plot of $\delta V_{\mathrm{pn}}$ against $N$ for all isotopic chains in this region. The \textsubscript{60}Nd and \textsubscript{72}Hf chains are drawn in black, or red where data are from this work, with species discussed in the text labelled. All other chains are shown in light grey for reference. Theoretical predictions are overlaid for the \textsubscript{72}Hf chain, from the BSkG03 model~\cite{Grams2023} and this work. \textbf{Bottom:} A 3-dimensional heat-map of $\delta V_{\mathrm{pn}}$ against $N$ and $Z$. New values from this work are indicated with a red box. The solid line corresponds to equal numbers of valence protons and valence neutrons; the dashed line corresponds to equal numbers of proton holes and neutron holes. Dotted mid-shell lines at $N=104$ and $Z=66$ demarcate the particle-particle, particle-hole, hole-hole, and hole-particle quadrants (from bottom-left clockwise). Elemental maxima according to the existing literature are marked with white circles from \textsubscript{52}Te to \textsubscript{74}W; this highlights the known effect along the line of nuclei with approximately equal numbers of valence protons and valence neutrons. The newly identified maximum in the hafnium chain, from this work, is marked with a white diamond; this highlights the new effect identified in this work, in a nucleus with approximately equal numbers of proton holes and neutron holes.}}
    \label{fig:Vpn1}
\end{figure}

The interdependence of the binding energies of the protons and neutrons can be investigated through the $\delta V_{\mathrm{pn}}$, a double-difference of four binding energies formulated to isolate the interaction of the final two protons with the final two neutrons~\cite{Zhang1989}. For even-even nuclei, the $\delta V_{\mathrm{pn}}$ are calculable as either of:

\begin{align}
\delta V_{\mathrm{pn}}(N,Z)
&= \frac{1}{4}\Bigl\{ S_{2\mathrm{p}}(N,Z) - S_{2\mathrm{p}}(N-2,Z) \Bigr\} \nonumber \\
&= \frac{1}{4}\Bigl\{ S_{2\mathrm{n}}(N,Z) - S_{2\mathrm{n}}(N,Z-2) \Bigr\}.
\label{eq:dvpn}
\end{align}

where the $S_{2\mathrm{p}}$ are analogous two-proton separation energies. Higher $\delta V_{\mathrm{pn}}$ values indicate stronger mutual binding between the last two protons and last two neutrons; the final neutron pair is more bound due to the presence of the final proton pair, and vice-versa. Six new ytterbium measurements unlock three new even-even $\delta V_{\mathrm{pn}}$ values for hafnium, utilising the relevant ytterbium $S_{2\mathrm{n}}$ inputs. From Figure~\ref{fig:Vpn1} (top) we can identify a maximum at $N=114$, associated with the divergence of the ytterbium and hafnium $S_{2\mathrm{n}}$. This new maximum is unexpected. Unlike prior elemental maxima above \textsuperscript{132}Sn -- including the $N=106$ peak in hafnium, which was the highest value in that chain prior to this work -- the relevant nucleus does not have an approximately equal number of valence protons and valence neutrons~\cite{Cakirli2025}. This previously identified trend, highlighted with a solid line in Figure~\ref{fig:Vpn1} (bottom), has been understood as the consequence of the strong alignment of similar, sequentially populated proton and neutron particle orbits~\cite{Cakirli2010}; this mechanism does not explain what appears to be a new effect at $N=114$ in hafnium. Moreover, from Figure~\ref{fig:Vpn1} (top) it can be seen that the only peak of a similar scale above \textsuperscript{132}Sn is at \textsuperscript{152}Nd; this is a known anomaly, occurring at the $N=90$ spherical-to-prolate transition~\cite{Brenner2006}.

\begin{figure}
    \includegraphics[width=\linewidth]{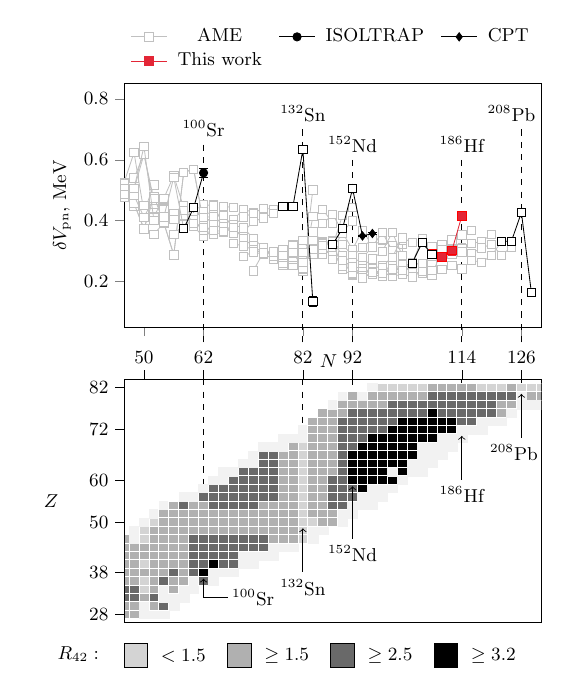}
    \caption{\textit{Experimental $\delta V_{\mathrm{pn}}$ (top) and $R_{42}$ (bottom) between \textsuperscript{76}Ni and \textsuperscript{210}Pb. Notable species discussed in the text are labelled, with the relevant sections of the $\delta V_{\mathrm{pn}}$ chains drawn in black. Data from the AME~\cite{Wang2021}, NNDC~\cite{NNDC}, CPT~\cite{Orford2022}, ISOLTRAP~\cite{Lunney2025}, and this work.}}
    \label{fig:E4E2}
\end{figure}

In Figure~\ref{fig:E4E2} (top), $\delta V_{\mathrm{pn}}$ values are shown from \textsuperscript{74}Ni to \textsuperscript{212}Pb. In Figure~\ref{fig:E4E2} (bottom), $R_{42} = E_{4+}/E_{2+}$ values highlight trends in deformation. This ratio of the lowest lying $4^{+}$ and $2^{+}$ excitation energies can provide an indirect view of trends in deformation, as it tends towards $R_{42}=3.33$ for an ideal rigid rotor and $R_{42}=2.0$ for near-spherical vibrational nuclei. This is due to the $J(J+1)$ band structure of low-lying rotational excitations in non-spherical deformed nuclei, and the approximately equal spacing of the lowest-lying one-phonon and two-phonon excitation modes of near-spherical, vibrational nuclei. Accounting for the decreasing macroscopic trend of the $\delta V_{\mathrm{pn}}$, one can see from Figure~\ref{fig:E4E2} (top) that the \textsuperscript{186}Hf maximum is similar in scale to peaks at  \textsuperscript{100}Sr, \textsuperscript{132}Sn, \textsuperscript{152}Nd, and \textsuperscript{208}Pb. In each of these cases a qualitative structural effect can be associated with the unusually strong dependence of the binding of the final two neutrons on the presence of the final two protons -- either double-magicity, as at \textsuperscript{132}Sn and \textsuperscript{208}Pb, or the sudden onset of deformation, as at \textsuperscript{100}Sr and \textsuperscript{152}Nd. The effect at \textsuperscript{186}Hf is comparable in magnitude to both examples, highlighting an urgent need for additional mass and spectroscopic data to to better understand this new effect.

Without spectroscopic data, we can leverage mean-field nuclear model descriptions to explore our current understanding. In Figure~\ref{fig:s2n} and Figure~\ref{fig:Vpn1} respectively we compare the $S_{2\mathrm{n}}$ and $\delta V_{\mathrm{pn}}$ to blind predictions from the Brussels-Skyrme BSkG03 dataset~\cite{Grams2023} and a bespoke Hartree-Fock plus Bardeen-Cooper-Schrieffer (BCS) approach with a Skyrme interaction~\cite{Bonneau2015}. The latter utilises the SIII parameterisation of the Skyrme interaction with time-odd terms and self-consistent blocking applied to odd-mass nuclei, as detailed in~\cite{Bonneau2015,Spataru25,Minkov2024}. These HF-BCS-Sk predictions are `blind' in that they were tuned to the pre-existing data, incorporating no new information from this work. The BCS pairing constants were determined through smooth variation to approximately reproduce the $E_{2+}/6$ moment of inertia factor of the $2^{+}$ states in the ground bands of the even-even nuclei, as in~\cite{Spataru25}. Details on the determination of the pairing strengths with respect to the moment of inertia for a particular nucleus are given in~\cite{Minkov2024}. An approximately parabolic trend with $N$ was followed for the $2^{+}$ states, as in Figure 17 of~\cite{Zyriliou2022}. This reflects in a corresponding behaviour of the pairing constants with a minimum in the best rotors region. In ytterbium, this approach is complicated by the distinct $\beta_{2}$ maximum and $E_{2+}$ minimum that exist at $N=102$ and $N=104$ respectively, and so two corresponding sets of values were obtained (`HF-BCS-Sk-1' and `HF-BCS-Sk-2') to take the two minima into account separately via two sets of results. The degree of similarity indicates that the model is robust against this choice. For hafnium only one set of values was obtained, `HF-BCS-Sk-1'.

For the $S_{2\mathrm{n}}$, shown in Figure~\ref{fig:s2n}, both the HF-BCS-Sk and BSkG03 models perform well. HF-BCS-Sk accurately predicts the macroscopic trend from \textsuperscript{178}Yb to \textsuperscript{184}Yb, to within \SI{200}{\kilo\electronvolt} for the most exotic four isotopes. The known flattening towards \textsuperscript{178}Yb appears to be reproduced by HF-BCS-Sk one neutron early, but is not captured by BSkG03. The steep descent from \textsuperscript{182}Yb to \textsuperscript{184}Yb is captured but underestimated by both models. As the known flattening towards \textsuperscript{186}Hf is well described by  HF-BCS-Sk, one can see from Figure~\ref{fig:Vpn1} that this results in a high $\delta V_{\mathrm{pn}}$ value at $N=114$. However, rather than an exceptionally sudden increase from $N=112$ to $N=114$, HF-BCS-Sk appears to predict a relatively sudden increase from $N=110$ to $N=112$, and thereafter a gradual increase to a maximum at $N=116$. With BSkG03 no flattening of the $S_{2\mathrm{n}}$ is apparent in hafnium until \textsuperscript{187}Hf, prior to a dramatic increase in neutron binding -- from Figure~\ref{fig:Vpn1} one can see that this results in a sudden increase in the $\delta V_{\mathrm{pn}}$ at $N=116$, two neutrons past $N=114$. The previously identified $\delta V_{\mathrm{pn}}$ peak at $N=106$ in \textsuperscript{178}Hf is not evident with BSkG03, but is captured by HF-BCS-Sk.

Within the BSkG03 model, the $S_{2\mathrm{n}}$ flattening in hafnium towards $N=115$ (Figure~\ref{fig:s2n}), and subsequent increase in binding for the final two neutrons at $N=116$ (relative to $N=114$), is associated with a sudden transition from a prolate $N=114$ ground state to an oblate $N=115$ ground state. This is evident from the $\beta_2$ deformation parameter, shown in Figure~\ref{fig:s2n} (bottom). This parameter is extracted from observables including the $B(E2)$ electric quadrupole transition probability and the spectroscopic quadrupole moment. It quantifies the quadrupolar component of the nuclear shape deformation, taking a positive or negative sign for a prolate or oblate shape respectively. The BSkG03 predictions suggest that the $\delta V_{\mathrm{pn}}$ peak predicted at $N=116$ may be driven by an earlier and more rapid increase in oblate deformation energy in hafnium, compared to ytterbium. This highlights one potential explanation for the very high $\delta V_\mathrm{pn}$ value observed experimentally. However, the HF-BCS-Sk model, which also predicts a relatively sudden increase in $\delta V_{\mathrm{pn}}$ in hafnium, predicts a much later transition at $N=118\text{-}120$, with an approximately \SI{200}{\kilo\electronvolt} deeper prolate minimum at $N=118$ and an approximately \SI{700}{\kilo\electronvolt} deeper oblate minimum at $N=120$. The effect of this on $S_{2\mathrm{n}}$ is apparent in Figure~\ref{fig:s2n}. This later transition aligns with predictions from the proxy-SU(3) algebraic model~\cite{Bonatsos2017}. The discrepancy between the HF-BCS-Sk and BSkG03 models, and their inability to accurately reproduce the experimental signal, underscores the need for additional mass and spectroscopic data.

As highlighted in Figure~\ref{fig:Vpn1} (bottom), the nuclei above \textsuperscript{132}Sn exhibit a striking `mini-valence Wigner' effect, with elemental $\delta V_{\mathrm{pn}}$ maxima consistently aligned with the nuclei with approximately equal numbers of valence protons and valence neutrons~\cite{Cakirli2025}. From particle-hole symmetry arguments, it seems plausible that a similar effect may occur in nuclei with approximately equal numbers of proton holes and neutron holes. However, the dominant explanation for the particle valence effect points to the overall similarity of the lower orbit sequences for the protons and neutrons in the relevant major shells~\cite{Cakirli2010}. As the higher orbit sequences for the protons and neutrons are dissimilar, with very different quantum numbers, this may preclude any expectation of a symmetric hole valence effect. The most aligned orbits would be expected to follow a less regular pattern, leading to a less regular trend in the $\delta V_{\mathrm{pn}}$.

This new maximum at $N=114$ is at \textsuperscript{186}Hf, a nucleus which does not have approximately equal numbers of valence protons and valence neutrons (with $N_{\mathrm{val}}=32$ and $Z_{\mathrm{val}}=22$) but does have approximately equal numbers of proton holes and neutron holes (with $N_{\overline{\mathrm{val}}}=12$ and $Z_{\overline{\mathrm{val}}}=10$). This is highlighted in Figure~\ref{fig:Vpn1}, where the dashed line corresponds to equal numbers of proton holes and neutron holes. This is an unexpected result, arising from the first direct $\delta V_{\mathrm{pn}}$ measurement within two neutrons of this line, across the entire major shell.

If this represents the emergence of a symmetric hole valence effect, in a region where the application of particle-hole symmetry is being explored, then further investigation will be of significant interest to theorists (see the application of contemporary ab initio theory to particle valence spaces above closed shells~\cite{PhysRevLett.118.032502}, and now to very heavy nuclei~\cite{Hu2022}; this approach may soon be replicated for hole valence spaces below \textsuperscript{208}Pb). Conversely, this effect may be idiosyncratic in hafnium, driven by the specific single particle structure. Further investigation would then present an opportunity to better understand the microscopic nature of the $\delta V_{\mathrm{pn}}$ interaction, which may be critical to a description of the anticipated prolate-to-oblate transition. For this point, see for example the favoured description of the spherical-to-prolate transition as a consequence of aligned and strongly interacting proton and neutron orbits~\cite{Federman1977}. Further mass data can clarify the extent to which there is a correlation of the strength of the $\delta V_\mathrm{pn}$ with the line of equal numbers of proton holes and neutron holes, alongside theoretical and spectroscopic studies to investigate the underlying orbits.

\begin{figure}
    \centering
    \includegraphics[width=\linewidth]{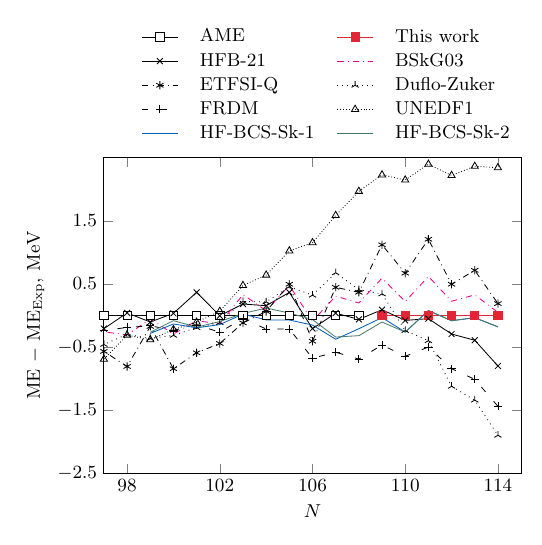}
    \caption{\textit{Ytterbium mass excesses from various theoretical models~\cite{Duflo1995,Pearson1996,Goriely2010,Moller2016,Grams2023}, including blind HF-BCS-Sk predictions from this work, presented as differences from the experimental values. Experimental values are from the AME~\cite{Wang2021} and first time mass measurements from this work. For the HF-BCS-Sk model the theoretical values plotted are obtained through a systematic correction of $\SI{-5}{\mega\electronvolt}$; see the text for details.}
    \label{fig:ME_comparison_new}}
\end{figure}

As a final aspect of this work, we plot the new experimental ground-state mass excesses alongside various predictions in Figure~\ref{fig:ME_comparison_new}, to benchmark the performance of these models towards $N=126$, a waiting point for the \textit{r}-process~\cite{Mumpower2016,Sun2008}. More accurate predictions for the neutron-rich rare-earths towards $N=126$ are essential to improve our understanding of the \textit{r}-process, to identify where the heaviest elements are created. Certain unexplained patterns in the relevant abundance curves are expected to arise from unknown structural effects in this region. In Figure~\ref{fig:ME_comparison_new}, we include predictions from the HFB-21, ETFSI-Q, Duflo-Zuker, FRDM, and BSkG03 models~\cite{Duflo1995,Pearson1996,Goriely2010,Moller2016,Grams2023}, and the new HF-BCS-Sk values. For HF-BCS-Sk the theoretical values plotted in Figure~\ref{fig:ME_comparison_new} are obtained through a systematic correction of $\SI{-5}{\mega\electronvolt}$ to obtain overall agreement with the previously measured AME2020 values. This facilitates a more equivalent comparison between the predictive performance of this model, which has been tuned to describe relative effects within isotopic chains, with the other models shown, which have already been tuned to the absolute scale of the mass surface. This correction does not affect the $S_{2\mathrm{n}}$ and $\delta V_\mathrm{pn}$ predictions discussed above, as this offset cancels  for these differences of binding energies. The seven models diverge by over \SI{3}{\mega\electronvolt}, only five neutrons beyond the pre-existing data.

Of the existing models, BSkG03 and HFB-21 appear to perform best, respectively remaining within \SI{0.62}{\mega\electronvolt} and \SI{0.80}{\mega\electronvolt} of the experimental data, with respective standard errors of \SI{0.20}{\mega\electronvolt} and \SI{0.28}{\mega\electronvolt} for the new data. BSkG03 displays some underestimation of binding and a staggering effect, the latter characteristic of an overestimated pairing strength. ETFSI-Q shows similar, accentuated behaviour. HFB-21, FRDM and Duflo-Zuker progressively overestimate the binding. The new HF-BCS-Sk values capture the overall trend in binding well, while also showing staggering. This systematic overestimation of pairing is a recurring challenge for models in this region~\cite{Vilen2018}. These results provide an anchor point for future models, facilitating more accurate predictions in a region where new data has been scarce. This will facilitate more accurate simulation of the \textit{r}-process, towards clarifying the origin of the heavy elements.

In summary, we have presented six first-time mass measurements of neutron-rich ytterbium, facilitated by advances in the production of exotic rare-earth beams. These measurements identify an anomalously high $\delta V_{\mathrm{pn}}$ proton-neutron interaction in \textsuperscript{186}Hf, a nucleus with approximately equal numbers of proton holes and neutron holes. This effect, which is comparable in scale to similar signals at doubly-magic nuclei and shape transitions, is not accurately reproduced by the contemporary models applied in this work. This highlights a gap in our understanding of both the proton-neutron interaction and the local structural evolution, as we begin to access this collective hole-hole regime. By reaching much farther from stability on the neutron-rich side, these measurements provide an important benchmarking exercise and key anchor point for contemporary models, towards an \textit{r}-process waiting point. This work provides first insights into interesting physical effects, and opens the door to investigating the effects of the strong force in a previously inaccessible region. It thus provides strong motivation and underscores the urgent need for further investigations leveraging improved production at next-generation radioactive ion beam facilities, that can now access this region~\cite{Tarasov2024}.


\section{Methods}\label{Methods}

\subsection{Multiple-Reflection Time-of-flight Mass-Spectrometry}

Time-of-flight mass spectrometry determines
the mass-to-charge ratio, $m/q$, from the time-of-flight $t_\mathrm{tof}$ of ions over a fixed path length $l$ in an electrostatic field $U(l)$:
\begin{equation}
  \frac{m}{q}= \frac{2U(l)}{l^2} t_\mathrm{tof}^{2}.  
\end{equation}
The parameters in the fraction are typically constant and can thus be combined as a device-specific calibration parameter $a$, which then depends on the ion kinetic energy and path length. However, due to signal propagation and electronic delays, the measured time $t_\mathrm{exp}$ differs from the real time-of-flight $t_\mathrm{tof}$, leading to a calibration equation 
\begin{equation}
\frac{m}{q} = a (t_\mathrm{exp} - t_{0})^{2},      
\end{equation}
where $t_{0}$ is a setup-specific offset. In order to increase the resolving power of such a time-of-flight mass spectrometer, which refers to its ability to distinguish ions with very similar mass-to-charge ratios, the flight path can be extended by storing ions for multiple reflections between two electrostatic isochronous ion mirrors~\cite{WOLLNIK1990267}. Here `isochronous' means that ions of the same energy take equal time to travel, regardless of their path length. In the limit of a long flight path, as the number of turns $N$ goes to infinity, the achievable mass resolving power of this multiple-reflection time-of-flight mass spectrometer can be taken as independent of the initial peak width
\begin{equation}
 \lim\limits_{N \to \infty} R_{\mathrm{m}}= \frac{t_{\mathrm{IT}}}{2\Delta t_{\mathrm{IT}}},   
\end{equation}
where $\Delta t_\mathrm{IT}$ is the peak broadening per turn due to ion-optical aberrations. Further details on the operation and performance of the TITAN MR-ToF-MS can be found in~\cite{Reiter2021}. For this work, resolving powers of $R_{\mathrm{m}} = {m}/{\Delta m}\sim 4\times 10^{5}$ were achieved.

\subsection{Data analysis}

Mass values were obtained from the time-of-flight spectra according to the data analysis procedure described in~\cite{Ayet2019,Paul2021}. The time offset $t_{0}$ was determined from offline measurements using \textsuperscript{85,87}Rb\textsuperscript{+} and \textsuperscript{133}Cs\textsuperscript{+} ions. The calibration parameter $a$ was determined via the times-of-flight of known ions present in the cocktail RIB delivered to TITAN, as given in Table~\ref{tab:mass_table}. A time-resolved calibration (TRC) was performed through the TOFControl software~\cite{Dickel2019}. Where a suitable TRC calibrant was not available in the cleaned isobaric beam, \textsuperscript{133}Cs\textsuperscript{+} was injected with post re-trapping beam merging. Peak centroids were identified through fitting spectra with hyper-exponentially modified exponential Gaussian functions~\cite{Purushothaman2017}, using the open-source Python package emgfit~\cite{Paul2024}. Statistical and systematic uncertainties were assigned following~\cite{Ayet2019,Paul2021}. For this work, the largest contributions to the overall uncertainty arose from systematic effects due to the non-ideal mirror switching.

The \textsuperscript{182}Yb—\textsuperscript{163}Dy\textsuperscript{19}F and \textsuperscript{184}Yb measurements exhibit higher relative uncertainties of $\delta m/m = 2\cdot10^{-7}$ and $\delta m/m = 2.2\cdot10^{-7}$, respectively, due to lower statistics. The repeat \textsuperscript{182}Yb—\textsuperscript{182}W measurement was performed to secure additional statistics, with a distinct beam composition favouring the differing calibrant. However, due to space charge effects, a $\delta m/m = 2\cdot10^{-7}$ systematic uncertainty was incurred for this measurement, resulting in a higher relative uncertainty of $\delta m/m = 2.5\cdot10^{-7}$. Combination of these two \textsuperscript{182}Yb measurements according to the procedure described in~\cite{Ayet2019} returns an adjusted value with a reduced relative uncertainty of $\delta m/m = 1.8\cdot10^{-7}$. All other new Yb measurements were obtained with less than $\delta m/m < 1.7\cdot10^{-7}$ uncertainty.

To further validate the mass accuracy and the analysis procedure, the masses of contaminant species with well known literature masses were determined in parallel. In total the masses of $34$ additional isotopes were obtained. Figure~\ref{fig:deviations} shows the deviation from literature of these determinations of those well known masses, comparing to the Atomic Mass Evaluation~\cite{Wang2021}. A mean deviation of approximately $-\SI{1.2(1.3)}{\kilo\electronvolt}$ with a spread of $\sigma=\SI{24}{\kilo\electronvolt}$ and a Birge ratio of $0.98$ provided confidence in the accuracy and precision of the system and analysis procedure.

\addtocounter{figure}{1}
\begin{figure*}[!b]
    \centering
    \includegraphics[width=0.8\linewidth]{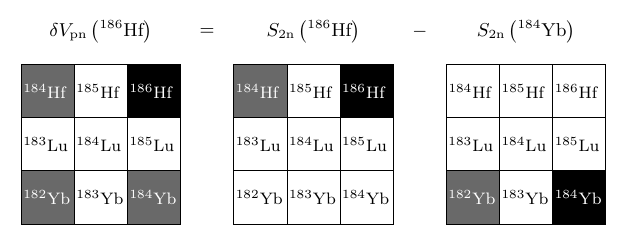}
    \caption{\textit{Graphical representation of the four isotopic mass inputs required for calculating the $\delta V_{\mathrm{pn}}$ value for \textsuperscript{186}Hf.}}
    \label{fig:DVpn_explain}
\end{figure*}
\addtocounter{figure}{-2}

For the calculation of second-order observables, such as $S_{2\mathrm{n}}$ and $\delta V_{\mathrm{pn}}$, additional isotopic mass values from the literature were used; uncertainties shown for these second-order observables reflect the combined uncertainties from the underlying data. Literature values were taken from both the Atomic Mass Evaluation and additional peer-reviewed publications, as indicated by the relevant citations in the main text. Figure~\ref{fig:DVpn_explain} illustrates the dependency of both $S_{2\mathrm{n}}$ and $\delta V_{\mathrm{pn}}$ on additional, neighbouring isotopic mass values. Although the valence proton–neutron interaction is attractive, implying $\delta V_{\mathrm{pn}}$ is physically negative (for most nuclei), we calculate $\delta V_{\mathrm{pn}}$ as typically positive -- from \eqref{eq:dvpn} -- and plot it accordingly, following the convention of virtually all previous studies.

\vfill
\begin{center}
\includegraphics[width=\linewidth]{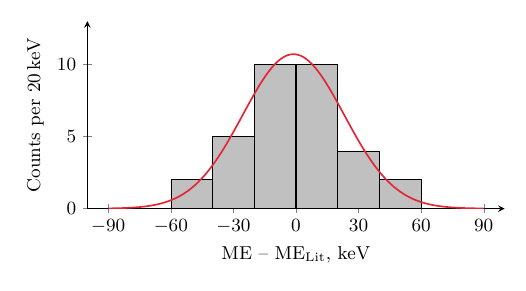}
\captionof{figure}{\textit{Distribution of deviations of well known species from literature~\cite{Wang2021}, contemporaneously measured as contaminants in the cocktail RIB. Known masses are reproduced with a mean deviation of $-\SI{1.2(1.3)}{\kilo\electronvolt}$, a spread of $\sigma=\SI{24}{\kilo\electronvolt}$, and a Birge ratio of $0.98$.}}
\label{fig:deviations}
\end{center}

\FloatBarrier


\section{Acknowledgements}

We would like to thank the Targets and Ion Source group at TRIUMF for developing the neutron-rich ytterbium secondary beam. This work was supported by the Natural Sciences and Engineering Research Council (NSERC) of Canada under Grants No. SAPIN-2018-00027, No. RGPAS-2018-522453, and No. SAPPJ-2018-00028; the National Research Council (NRC) of Canada through TRIUMF; the Canada-UK Foundation; the UK Science and Technology Facilities Council (STFC) under Grants No. ST/V001051/1 and No. ST/Y000293/1; German institutions DFG (grants FR 601/3-1, contract no.\ 422761894 and SFB 1245, and through the PRISMA Cluster of Excellence) and BMBF (grants 05P21RGFN1 and 05P24RG4), and by the JLU and GSI under the JLU-GSI strategic Helmholtz partnership agreement. N.M. acknowledges the support by the Bulgarian National Science Fund (BNSF) under Contract No. KP-06-N98/2.

\bibliography{sn-bibliography}

\end{document}